\documentstyle[prb,aps,epsf,psfig,multicol]{revtex}

\newcommand{\be}{\begin{equation}}
\newcommand{\ee}{\end{equation}}
\newcommand{\bea}{\begin{eqnarray}}
\newcommand{\eea}{\end{eqnarray}}

\begin{document}
\draft

\title{Decoupling of the $S=1/2$ antiferromagnetic zig-zag ladder
with anisotropy }  

\author{V. R. Vieira}
\address{Departamento de F\'{\i}sica and  CFIF, Instituto Superior 
T\'ecnico, Av. Rovisco Pais, 1049-001 Lisboa, Portugal}

\author{N. Guih\'{e}ry}
\address{Laboratoire de Physique Quantique, Universit\'{e} Paul
Sabatier, F-31062 Toulouse, France and \\
CFIF, Instituto Superior
T\'{e}cnico, Av. Rovisco Pais, 1049-001 Lisboa, Portugal}

\author{J. P. Rodriguez}
\address{
Department of Physics and Astronomy, California
State University, Los Angeles, California 90032, USA}

\author{P. D. Sacramento}
\address{Departamento de F\'{\i}sica and  CFIF, Instituto Superior 
T\'ecnico, Av. Rovisco Pais, 1049-001 Lisboa, Portugal}

\date{\today}

\maketitle

\begin{abstract}
The spin-1/2 antiferromagnetic
zig-zag ladder is studied by exact diagonalization of small systems 
in the regime of  weak inter-chain coupling.
A gapless phase with quasi long-range spiral correlations 
has been predicted to occur in this regime if easy-plane ($XY$)
anisotropy is present. 
We find in general that the finite zig-zag ladder
shows three   phases: a gapless 
collinear phase, a dimer phase and a spiral phase. 
We study the level crossings of the spectrum,
the dimer correlation function, the structure factor
and the spin stiffness within these phases, 
as well as at the transition points. 
As the inter-chain coupling decreases
we observe a transition in the anisotropic $XY$ case from a
phase with a gap to a gapless phase that is best described
by two decoupled antiferromagnetic chains.
The
isotropic and the  anisotropic $XY$ cases are found to be 
qualitatively the same, however, 
in the regime of weak inter-chain coupling for the  small  systems
studied here.  We attribute this to a finite-size effect
in the isotropic zig-zag case that results from 
exponentially diverging antiferromagnetic
correlations in the weak-coupling limit.
\end{abstract}
\vspace{0.3cm}
\pacs{PACS numbers:}
\begin{multicols}{2}
\section{Introduction}
Antiferromagnetic ladder systems have attracted 
much  interest recently.~\cite{Dagotto1}
On the theoretical side they
interpolate between the well studied antiferromagnetic chain~\cite{Haldane1}
and two dimensional antiferromagnets.~\cite{CHN}
The evolution between 1D and 2D spin-1/2 antiferromagnetism is not
necessarily smooth, however. 
In particular, the $n$-leg ladder shows a
remarkable alternating  property   in the spectrum as the number of legs
is even or odd.~\cite{Rice}  
The spectrum has a gap
for an even number of legs while it is gapless for an odd number of
legs.
This is similar to the difference between integer (spin-gap) 
and half-odd-integer
(gapless) spin chains~\cite{Haldane2}. 
In the limit of strong-coupling between
the two chains the two-leg ladder is essentially composed of 
weakly interacting singlets that form across the rungs.
The lowest excitation is the promotion
of a rung singlet to a triplet with an excitation
energy of the order of the
inter-chain coupling. 
This spin gap remains nonvanishing even for small inter-chain   coupling 
due to the fact that a single antiferromagnetic
chain is  critical.~\cite{Barnes} In the case of purely  Ising coupled
chains the gap appears for all values of $n$.~\cite{Rodriguez}

The  antiferromagnetic zigzag ladder has also attracted interest recently,  
particularly in the context of
experimental systems with low-dimensional
magnetic structures like that of
$Cs_2CuCl_4$~\cite{Coldea}.
It is also interesting from a theoretical point of view
because it is a frustrated system (see Fig. 1).
Indeed, the zig-zag ladder is 
equivalent to a single antiferromagnetic 
chain with next-nearest-neighbor interactions.
In this paper we consider the spin-1/2 antiferromagnetic zigzag ladder
with anisotropy.
The isotropic case has been studied 
before~\cite{Okamoto,Eggert,Nathalie,Chitra,White96,Allen}
as a function of the coupling parameter, $j = J_2/J_1$, which is the ratio
of the next-nearest-neighbor interaction, $J_2$, to the nearest-neighbor
interaction, $J_1$.
As $j$ increases, the system goes from gapless (single chain)
to a dimer phase and then to a spiral phase, where the structure factor
has a maximum at a momentum $\pi/2<q<\pi$. 
The system has a spin gap 
in these last  two phases,
and it  therefore only displays short-range order. 
In the limit that
the intra-chain interaction is much larger than the inter-chain interaction
($j \rightarrow \infty$) the two chains decouple and a gapless single chain
behavior is recovered. 
It has been argued that this  only happens, strictly speaking,
at $j= \infty$: the spin gap becomes exponentially 
small as $j$ grows, but it remains non-vanishing.~\cite{White96}
Recently, on the other hand,
 it has been proposed that
incommensurate quasi-long-range spin correlations should be observed
if easy-plane ($XY$)
anisotropy is included in the zigzag ladder.~\cite{Nersesyan} 
This is argued
to be due to the presence of a ``twist'' term
that results from the inter-chain
interaction. It has been proposed that there is one gapless mode and one mode
with a  gap in the regime of strong $XY$ anisotropy in the 
inter-chain coupling. Another prediction
of this work is the existence of
spontaneous local 
spin currents.  This,  however,
has been refuted in ref. 16. 
Also, other recent numerical  work~\cite{Aligia}
has 
failed to
confirm the gapless nature of the groundstate
in the anisotropic $XY$ case
at weak interchain coupling. 
Recent Density Matrix Renormalization Group (DMRG) 
results~\cite{Hikihara} suggest, however, that the zig-zag ladder 
does indeed show a gapless chiral
phase as predicted in ref. 15.

In this paper we use exact diagonalizations, the modified Lanczos 
method~\cite{Gagliano} and the Davidson method~\cite{Davidson} 
to address the possibility of a transition from a spin-gap regime
to a gapless regime as a function of $j$ when anisotropy is present
in small $S = 1/2$ antiferromagnetic zig-zag ladders.
We compute various  probes to identify 
the different phases and study their behavior
close to the transition points. 
We study in particular  the spin stiffness, the dimer correlation
function, the structure factor, 
and analyze in detail the spectrum 
in the various parameter regimes. 
Since the zig-zag ladder effectively  has both nearest-neighbor and 
second neighbor interactions, 
a stiffness tensor is required to account
for these two types of interactions. 
The eigenvalues
of this $2\times 2$ matrix then become  the natural spin
rigidities that we use
to clarify the behavior of the system in the various regimes. 
The stiffness of a system is a particularly  good
measure of the long-range nature of the groundstate. 
Introducing twisted boundary conditions
leads to a response in the energy if the quantum states 
are extended (gapless case). 
On the other hand, the energy is insensitive to a change
in the boundary conditions
if the quantum states are localized (spin-gap case). 
Therefore, the stiffness with respect to such a  twist is positive if the
system is gapless and it is zero if the system has a gap.
Also, the dimer correlation function naturally signals the dimer phase
while
the structure factor is a natural way to detect and study the spiral phase.

Our results are consistent with a gapless excitation
spectrum  in the case of  $XY$ anisotropy at
weak interchain coupling. 
We obtain qualitatively similar results
for the isotropic case, however.
This is most  likely a   finite size effect
due to the exponentially small spin gap  that persists
in the isotropic zig-zag at weak coupling
in the thermodynamic limit.~\cite{White96}
In fact, we show that the phase diagram that is  obtained from the analysis
of the spectrum for finite systems may be consistent with 
the field theory prediction~\cite{Nersesyan}
after performing extrapolations to the thermodynamic limit.

The paper is organized as follows: In section II we present the model and the
quantities to be calculated. In section III we present our results and in
section IV we summarize the work. 
Technical details concerning exact diagonalization 
are given in the Appendix.
We also briefly review the extrapolation
technique to the thermodynamic limit here.

\section{Model and Probes}

The anisotropic zigzag ladder is defined by the Hamiltonian
\bea
H & = & \frac{1}{2} J_{1}^{XY} \sum_{i} \left( S_{i}^{+} S_{i+1}^{-} +
S_{i}^{-} S_{i+1}^{+} \right) + J_1^{z} \sum_{i} S_{i}^{z} S_{i+1}^{z} \nonumber \\
& + & \frac{1}{2} J_{2}^{XY} \sum_{i} \left( S_{i}^{+} S_{i+2}^{-} +
S_{i}^{-} S_{i+2}^{+} \right) + J_2^{z} \sum_{i} S_{i}^{z} S_{i+2}^{z}.
\eea
The spin operators refer to spin $S = 1/2$ states, while
the summation $i=1,...,N$ runs along the ``rib'' of the zig-zag ladder.
We shall  parameterize the interactions by the 
coupling parameter $j=J_2^{XY}/J_1^{XY}$ and
by the anisotropy parameter
$J_1^z/J_1^{XY}=\Delta=J_2^z/J_2^{XY}$.
(The isotropic case reduces to $j=J_2/J_1$ and $\Delta=1$.) 
We will set $J_1^{XY}=1$ henceforth. 
Consider first the  nearest-neighbor Heisenberg
chain with anisotropy, which corresponds to both the weak-coupling
($J_1 = 0$) and to the strong-coupling ($J_2 = 0$) limits of the
zig-zag ladder (see Fig. 1). The spectrum is gapless for
the case of $XY$ anisotropy, $|\Delta| \leq 1$,
as shown by the Bethe ansatz.~\cite{Baxter} 
The excitation spectrum
consists of spin-$1/2$ particles dubbed spinons. 
Since flipping one spin 
represents a spin-$1$ excitation,
the spinons can only be created in pairs. 
Therefore the conventional
spin $1$ magnons are deconfined into 
spin-$1/2$ spinons that propagate incoherently.
In the regime where $\Delta \leq -1$, the groundstate is ferromagnetic.
When $-1\leq \Delta \leq 1$ the leading spin configuration is the N\'{e}el
state with the staggered magnetization lying within the $XY$ plane.
At $\Delta=1$ the groundstate
is again in a N\'{e}el state, but with a staggered magnetization that can point
in any direction. 
Last, the spectrum shows a  gap in the Ising regime at $\Delta>1$.
The groundstate, on the other hand,  displays strict long-range N\'{e}el order,
with the staggered magnetization directed along the $z$ axis.

We shall begin our study  of the antiferromagnetic zig-zag ladder
by analyzing the classical limit of  the isotropic Heisenberg
case first: $J_1^{XY}=J_{\perp}=J_1^{z}$ and $J_2^{XY}=J_{\parallel}=J_2^{z}$
as $S\rightarrow\infty$. 
A spiral state $S_i^{+} = S e^{i \theta_i}$ yields an energy per site of 
$E(\theta) = S^2 J_{\parallel} \cos (2 \theta) + S^2 J_{\perp} \cos \theta$.
This magnetic energy
is minimized at a pitch angle $\theta_0$  that satisfies 
$\cos \theta_0 = -\frac{1}{4}
J_{\perp}/J_{\parallel}$ for inter-chain exchange couplings that are below
 a critical value
$J_{\perp}^{c}= 4 J_{\parallel}$.
A ferromagnetic state on each chain
occurs, on the other hand, at strong coupling $J_{\perp}>J_{\perp}^{c}$,
with a pitch angle of $\theta_0 = \pi$.
The spins are thus arranged  antiparallel in between chains. 
To summarize,
the system is in a spiral phase for $J_{\perp} < 4 J_{\parallel}$,
while it is in a collinear phase for $J_{\perp}> 4 J_{\parallel}$. 
The same holds true 
when only $XY$-coupling exists. In the case of Ising coupling only,
on the other hand, we have the effective model 
$H=J_{\parallel}^{z} \sum_i S_i^z S_{i+2}^z + J_{\perp}^z \sum_i
S_i^z S_{i+1}^z$. 
There are two possible groundstates. The first is the collinear
state defined by $S_i^z= \frac{1}{2}$, for $i$ even and $S_i^z= - \frac{1}{2}$
for $i$ odd (this has a degeneracy $2$) with an energy per site of
$E=S^2 (J_{\parallel}^z - J_{\perp}^z)$. 
The other state is the antiferromagnetic
one defined by $\pm S_i^z = \frac{1}{2} (-1)^{i/2}$ 
for $i$ even and $\pm S_i^z = \frac{1}{2}
(-1)^{(i\pm 1)/2}$, for $i$ odd 
(this has a degeneracy $2 \times 2$), with an energy
per site of $E=-S^2 J_{\parallel}^z$. 
We have an Ising antiferromagnet
for $J_{\perp}^z < 2 J_{\parallel}^z$ and a collinear Ising ferromagnet
for $J_{\perp}^z > 2J_{\parallel}^z$,
with a first order transition separating the two phases.

Consider now the stiffness tensor in the classical limit.
Imposing a spiral spin configuration on the zig-zag ladder with a
pitch angle $\theta$, the energy per site of the classical $J_1-J_2$
model is then given as above by
\be
e=\frac{E}{N} = 
S^2 J_1 \cos (\theta + \theta_1) + S^2 J_2 \cos (2 \theta + \theta_2),
\ee
where we have added small twists $\theta_1$ 
and $\theta_2$ to the nearest-neighbor
and the next-nearest-neighbor terms, respectively.
For $\theta_1= 0 = \theta_2$, we have that $\cos \theta_0=-1$ for $J_1>4 J_2$
and that $\cos \theta_0 = -\frac{1}{4} (J_1 / J_2)$ 
for $J_1<4 J_2$, as stated
above. The spin currents are then given by
\bea
j_1 & = & \left. \frac{\partial e}{\partial \theta_1}\right|_0 
= -S^2 J_1 \sin \theta_0 \nonumber \\
j_2 & = & \left. \frac{\partial e}{\partial \theta_2}\right|_0 = -S^2 
J_2 \sin 2\theta_0
\eea
and the rigidity components by
\bea
\rho_{11} & = & \left. \frac{\partial j_1}{\partial \theta_1}\right|_0 
= -S^2 J_1 \cos \theta_0 
\nonumber \\
\rho_{12} & = & \left. \frac{\partial j_1}{\partial \theta_2}\right|_0 
= 0 \nonumber \\ 
\rho_{22} & = & \left. \frac{\partial j_2}{\partial \theta_2}\right|_0 
= -S^2 J_2 \cos 2\theta_0.
\eea
Note that both the spin currents and the stiffnesses are independent
of the anisotropy parameter $\Delta$
in the classical limit: the isotropic and the $XY$ anisotropic
cases give the same results. 
In the collinear phase at $J_1 > 4J_2$,
the spin currents vanish
($j_1= 0 = j_2$) and 
$\rho_{11}= S^2 J_1$, $\rho_{12}=0$ and $\rho_{22}=-S^2 J_2$. 
On the other hand, 
in the spiral phase at $J_1 < 4J_2$ the local spin currents are
non-vanishing:
$j_1=\pm S^2 J_1 [1-(J_1 / 4 J_2)^2]^{1/2}$ 
and $j_2=-\frac{1}{2} j_1$.
However, the total spin current $j_s=j_1 +2 j_2$ is null. 
The stiffness tensor
of this spiral phase 
is  given by 
$\rho_{11}=\frac{1}{4} S^2 (J_1^2 / J_2)$, $\rho_{12}=0$ and
$\rho_{22}=S^2 [J_2 - \frac{1}{8} (J_1^2 / J_2)]$. The natural stiffness
associated with the total spin $j_s$ is the
response to an external twist that satisfies $\theta_2=2 \theta_1$, 
 and is given by
$\rho_s=\rho_{11} + 4 \rho_{12} +4 \rho_{22}$. 
It reduces to $\rho_s=S^2 (J_1-4J_2)$ in the collinear phase 
 and to 
$\rho_s = S^2 [4J_2 - \frac{1}{4} (J_1^2 / J_2)]$ 
in the spiral phase. These results are displayed in Fig. 2. 
Here it is
shown that $\rho_s$ is always positive and 
that vanishes at the classical transition
point between the collinear ferromagnet to the spiral phase. Recall that
the nature of the groundstate changes across this transition. Also, 
we remark that in the spiral
phase only the stiffness $\rho_s$ and the 
total current $j_s$ are ``well behaved''.
The other components show spontaneous spin currents,
while  $\rho_{22}$ is
not always positive (we will return to this point later while discussing
the quantum case).  

In the general quantum case we calculate the stiffness in the 
standard way.~\cite{Shastry} 
We consider the Hamiltonian (1)
with periodic boundary conditions imposing uniform twists around the $z$
axis:
\bea
H(\theta_1, \theta_2) & 
= & \frac{1}{2} J_1^{XY} \sum_i \left(S_i^{+} S_{i+1}^{-}
e^{i \theta_1} + S_i^{-} S_{i+1}^{+} e^{-i \theta_1} \right) \nonumber \\
& + & \frac{1}{2} J_2^{XY} \sum_i \left(S_i^{+} S_{i+2}^{-}
e^{i \theta_2} + S_i^{-} S_{i+2}^{+} e^{-i \theta_2} \right) \nonumber \\
& + & J_i^z \sum_i S_i^z S_{i+1}^z + J_2^z \sum_i S_i^z S_{i+2}^z.
\eea
Here $\theta_1$ and $\theta_2$ are two independent twists that act separately
along the interchain and intrachain directions, respectively. Expanding the 
exponentials to second order we obtain the form
\bea
H(\theta_1, \theta_2) & 
= & H(0,0) + \theta_1 \sum_i J_i^1 + \theta_2 \sum_i J_i^2
\nonumber \\
& - & \frac{1}{2} \theta_1^2 \sum_i T_i^1 -\frac{1}{2} \theta_2^2 \sum_i T_i^2
\eea
where
\bea
J_i^1 & = & \frac{i}{2} J_1^{XY} 
\left( S_i^{+} S_{i+1}^{-} - S_i^{-} S_{i+1}^{+} \right)
\nonumber \\
J_i^2 & = & \frac{i}{2} J_2^{XY} 
\left( S_i^{+} S_{i+2}^{-} - S_i^{-} S_{i+2}^{+} \right)
\eea
are the spin currents along the interchain and intrachain directions, 
respectively and
where
\bea
T_i^1 & = & \frac{1}{2} J_1^{XY} 
\left( S_i^{+} S_{i+1}^{-} + S_i^{-} S_{i+1}^{+} \right)
\nonumber \\
T_i^2 & = & \frac{1}{2} J_2^{XY} 
\left( S_i^{+} S_{i+2}^{-} + S_i^{-} S_{i+2}^{+} \right)
\eea
are the kinetic energy  operators. Using second-order perturbation 
theory, we then obtain that
the groundstate stiffness tensor, 
$\rho_{\alpha \beta} =
(\partial^2 E /	\partial \theta_{\alpha} \partial \theta_{\beta})|_0$,
is given by
\bea
\rho_{\alpha \beta} 
= & - &\sum_i <0|T_i^{\alpha} |0> \delta_{\alpha,\beta} \nonumber \\
  & - & 2 \sum_{\nu \neq 0} \frac{1}{E_{\nu} - E_0}
\sum_{i,j} <0|j_{i}^{\alpha} | \nu> < \nu|j_{j}^{\beta} |0>,
\eea
where $|0>$ is the groundstate and $|\nu>$ are the excited states
($\alpha, \beta=1,2$).
The groundstate is assumed to be non-degenerate. 
Both the spin current operator 
$J^{\alpha}=\sum_i J_i^{\alpha}$ and the kinetic
operator $T^{\alpha}=\sum_i T_i^{\alpha}$ commute with 
the translation operator, $T$, and conserve
total spin $S_z$ (where $\alpha=1,2$). 
Therefore, the states $|\nu>$ in the stiffness 
formula (9) are
the excited states within the subspace of a given
magnetization and momentum
that contains the groundstate. 

As in the classical case discussed earlier,
the stiffness can also be calculated
directly~\cite{Bonca} taking numerical derivatives of the 
energy with respect to
small twists, $\theta_1$ and $\theta_2$, that are
imposed on the system. This procedure
requires care that $\theta_1$ and $\theta_2$ 
are small enough so that there be no
level crossings.  
(We prefer to use the 
correlation function method, even
though the results using both methods agree very well.)
The  change in energy due to small twists
$\theta_1$ and $\theta_2$ takes the form
\be
\delta E = \frac{1}{2} \rho_{11} \theta_1^2 
+ \rho_{12} \theta_1 \theta_2
+ \frac{1}{2} \rho_{22} \theta_2^2 
\ee
in the absence of spontaneous spin currents.
It is then natural to consider  the eigenvalues
of the stiffness tensor
\be
\rho_{\pm} = \frac{1}{2} 
\left( \rho_{11} + \rho_{22} \right) \pm \sqrt{ \left( \frac{\rho_{11}
- \rho_{22}}{2} \right)^2 + \rho_{12}^{2}}
\ee
and the determinant
$D=\rho_+ \rho_- = \rho_{11} \rho_{22} - \rho_{12}^{2}$. 
These eigenvalues will be computed using expression (9)
for the stiffness tensor in the next section.

We shall also calculate the correlation function associated with 
the dimerization 
via linear response theory.
Imposing a small explicit dimerization, we consider the
Hamiltonian
\bea
H(\delta_1, \delta_2) & 
= & \frac{1}{2} J_1^{XY} \sum_i \left( 1+ (-i)^i \delta_1
\right) \left( S_i^{+} S_{i+1}^{-} + S_i^{-} S_{i+1}^{+} \right) \nonumber \\
& + & \frac{1}{2} J_2^{XY} \sum_i \left( 1+ (-i)^i \delta_2
\right) \left( S_i^{+} S_{i+2}^{-} + S_i^{-} S_{i+2}^{+} \right) \nonumber \\
& + & J_1^z \sum_i \left( 1+ (-i)^i \delta_1
\right) S_i^{z} S_{i+1}^{z} \nonumber \\
& + & J_2^z \sum_i \left( 1+ (-i)^i \delta_2
\right) S_i^{z} S_{i+2}^{z}.
\eea
Once again using second order perturbation theory we obtain that
the susceptibility 
$\chi_{\alpha \beta} =
- (\partial^2 E / \partial \delta_{\alpha} \partial \delta_{\beta})|_0$ 
is given by
\bea
\chi_{\alpha \beta} &&= 
 2\,  {\rm Re} \sum_{\nu \neq 0} \frac{1}{E_{\nu} - E_0} \sum_{i,j} 
(-1)^{i+j} \nonumber \\
&&  <0|(T_{i}^{\alpha}+ M_{i}^{\alpha}) | \nu>
< \nu |(T_{j}^{\beta}+ M_{j}^{\beta})|0> 
\nonumber \\
&&
\eea
where
\bea
M_{i}^{1} &=& J_{1}^{z} S_{i}^{z} S_{i+1}^{z} \nonumber \\
M_{i}^{2} &=& J_{2}^{z} S_{i}^{z} S_{i+2}^{z}.
\eea
In the dimer correlation function, the factor
$(-1)^{i+j}$ implies that the states $|\nu>$ are 
contained in the same magnetization
subspace, but in the $k=k_0 \pm \pi$ momentum subspace, 
where $k_0$ is the groundstate
momentum.
Last, the structure factor is defined in the usual way:
\be
S(q) = \sum_{r} e^{i qr} C_r 
\ee
where the correlation function $C_r$ is defined by
\be
C_r = \frac{1}{N S(S+1)} \sum_i <0| \vec{S}_{i} . \vec{S}_{i+r} |0>.
\ee
and is normalized such that the local correlation function ($r=0$)
is unity.

\section{Results}
We now proceed to study the $XXZ$ zig-zag model (1) using exact diagonalization of finite systems
with sizes up to $N=28$. 
The full energy spectrum is obtained
for the smaller system sizes, $N \leq 16$,
while only the groundstate and
the first few excited 
states can be determined for the larger system sizes. 
The eigenvectors and eigenvalues of Hamiltonian eq. (1) are then
substituted into Eqs. (9,13,15) to compute the various correlation functions.
Our main aim is to  study the transitions between the various phases. 

{\it Spectrum.}  
Let us  first survey the energy spectrum that is displayed by these small
systems.  We shall keep track of important quantum numbers associated
with each energy  level, such as the momentum along the rib of 
the zig-zag, the
spin and the parity.  We shall also identify points in parameter
space where low-lying levels cross, and use this to identify
phase transitions in the system.
This procedure is known to yield accurate transition points when 
applied to even relatively small systems.~\cite{Okamoto,Tonegawa92} 

Let us begin by determining the  quantum numbers of the groundstate
as a function of the size $N$
for the $S = 1/2$ zig-zag antiferromagnet.
Periodic boundary conditions are imposed throughout.
The groundstate is 
a spin singlet in general due to the antiferromagnetic
interactions.
For strong enough coupling between chains, $j = J_2/J_1 < 1/2$, 
it
has either momentum $\pi$ for $N=4n+2$
or  momentum $0$ for $N=4n$. 
For weak enough coupling between chains $j>1/2$, on the other hand,
the momentum oscillates between $0$
and $\pi$ as a function of the coupling 
parameter $j$ and of the system size $N$.~\cite{Tonegawa87} 
There are several points along $j$ 
in this regime
where the corresponding  energy levels
for these two momentum values cross. 
The groundstate is degenerate at these points,
and this is reflected by  peaks in the 
dimer correlation function (see Fig. 8). 
Such level crossings grow in number as the system size grows,
and this indicates that the two singlet states in question
are in fact degenerate in the
thermodynamic limit.  
By the Lieb-Schultz-Mattis theorem,~\cite{Lieb} 
this is consistent with a 
spin gap in the excitation spectrum that survives the
thermodynamic limit  in the weak-coupling regime $j > 1/2$. 

Consider now the  specific case 
of an $XXZ$  zig-zag chain (1) with $N=16$ sites under    periodic
boundary conditions (see Table I). 
In the isotropic case, $\Delta = 1$,
the states are organized into spin multiplets due 
to the $SU(2)$ spin invariance.
Again, the  antiferromagnetic interactions
imply that  the groundstate is a spin singlet ($S_z = 0$)
in general.
The system has three well defined regimes: ({\it a}) strong-coupling,
({\it b}) intermediate coupling and ({\it c}) weak-coupling.
In the strong-coupling  limit ({\it a}),
$j = J_2/J_1\rightarrow 0$, 
the groundstate has momentum $k=0$. 
The first excited state forms a spin triplet in such case,
with $k=\pi$, while  the second excited state is	
another spin singlet with  $S_z=0$ and  $k=\pi$. 
At $j=j^N_{c1} \sim 0.24$ there is a level crossing where the first excited states
and the second excited state interchange (here $j^N_{c1}$ is the value of 
$j$ at which the level crossing occurs for the system with size $N$).
In the thermodynamic limit the two lowest singlet states
($S_z=0$, $k=0$ and $S_z=0$, $k=\pi$) become degenerate and there is
a finite gap to the next excited
state (the triplet state $S_z=0,\pm 1$, $k=\pi$).
Although the system begins to
dimerize at this stage ({\it b}),
antiferromagnetic correlations along the rib
of the zig-zag remain dominant up
to the Majumdar-Ghosh point~\cite{Majumdar}
at  $j=1/2$. 
The groundstate is doubly 
degenerate for any system size at this point, where the two
states are perfectly dimerized. 
Antiferromagnetic correlations {\it within} chains of the zig-zag
become dominant
beyond this point at weaker couplings $j > 1/2$ (see Fig. 9).
Another level crossing occurs
as $j$ increases to about $j^N_{c2} \sim 1.6$
such that the first excited
states  become
two triplets with $S_z=0,\pm 1$, $k=\pm \pi/2$.
The groundstate displays quantum numbers
$S_z=0$ and $k=0$ at this stage ({\it c}), 
and it is no longer degenerate.
This remains so as $j \rightarrow \infty$.
It should be mentioned,
however, that  White and Affleck predict
that a   non-zero spin gap persists in the thermodynamic
limit at  weak-coupling $j\gg 1$ between chains, 
and that it becomes exponentially small as
$j$ grows~\cite{White96} (we will return to this point later).

The spectrum of
the anisotropic $S = 1/2$ $XXZ$ zig-zag ladder has also been studied
previously 
in the strong-coupling regime
up to the Majumdar-Ghosh line ($0 < j < 1/2$).~\cite{Nomura}
A gapless regime occurs for $XY$ anisotropy  $\Delta \le 1$ and
strong coupling
$j<j_{c1}(\Delta)$,  an Ising antiferromagnet along the rib
of the zig-zag that shows a spin gap in the excitation spectrum
occurs for $\Delta>1$ and $j<j_{c1}(\Delta)$,
and a dimer phase regime that also has a spin gap
exists at  $j>j_{c1}(\Delta)$ and any $\Delta$
(here $j_{c1}(\Delta)$ is the transition line obtained from extrapolation
to the thermodynamic limit).
The line $j = j_{c1}(\Delta)$
separates the gapless phase from the dimer phase for $\Delta \le 1$,
while it
separates the dimer phase from the (Ising) N\'{e}el phase for $\Delta>1$.
The line at $\Delta=1$ and $j<j_{c1}$ separates the 
$XY$ gapless phase from the Ising phase.

Consider again the specific case of $N=16$ sites in this instance, 
with anisotropy parameters
 $\Delta=0$ or $\Delta=0.5$ (see Table I). 
The groundstate is a singlet with $S_z=0$ and $k=0$
as $j$ increases from the strong coupling limit at $j = 0$ up to $j=1/2$.
The first excited state has  degeneracy $2$, 
with spin   $S_z = \pm 1$ and momentum $k=\pi$
inside this regime. 
The next excited state is again a singlet with $S_z=0$ and $k=\pi$. 
As $j$ increases there exists a line of points, $j = j^N_{c1}(\Delta)$,
such that the energy level of the
 excited state with $S_z=\pm 1$, $k=\pi$ 
crosses the energy level of the
excited state $S_z=0$, $k=\pi$.~\cite{Nomura} 
For $j>j^N_{c1}(\Delta)$ the two lowest states are the two singlets
with $S_z=0$, $k=0$ and $S_z=0$, $k=\pi$. 
Again, 
these two states become degenerate (with a gap to the first excited state)
in the thermodynamic limit.  
The groundstate must be a spin singlet with $S_z = 0$
due to  the  antiferromagnetic couplings.
This  excludes the possibility that any  $S_z = \pm 1$
state be degenerate with it.
As a result,
the level crossing between the $S_z = \pm 1$ and singlet states
can be used as an accurate criterion to determine
the phase transition between the gapless and the spin-gap
regimes.~\cite{Okamoto}
If we  increase $j$ up to $j=1/2$, then  the system is exactly degenerate
for all system sizes.
This is a feature of the Majumdar-Ghosh point,
which has a perfectly  dimerized groundstate.
The behavior of the system does not change much
for intermediate coupling $j > 1/2$ beyond  this point, 
with the exception that
the momentum of the two lowest states interchanges between 
$k=0$ and $k=\pi$ as $j$ grows. 
At a larger value of $j=j^N_{c2}$ between $1.2-1.6$, however,
a level crossing occurs
between the $S_z=0$, $k=0\quad {\rm or}\quad \pi$ state (first excited state) 
and a state with $S_z=\pm 1$,
$k=\pi/2$ (second excited state). 
Notice that the  momentum of the first excited state
 is now $k=\pi/2$.
This  is to be expected in this regime
since the two chains are weakly coupled, and the periodicity doubles. 
For $j>2$, in particular,
the first excited state is now four-fold degenerate 
($S_z=\pm 1$, $k=\pm \pi/2$),
and we might expect to fall back into a gapless
regime since the first excited state is not a spin singlet. 
This level structure is in fact the same as for $j\rightarrow \infty$. 

In Fig. 3 we present the crossing points between 
the states $S_z=0$, $k=\pi$ and
$S_z=\pm 1$, $k=\pm \pi/2$ that signal this transition
as function of the number of sites
for several values of the anisotropy parameter $\Delta$. 
We have attempted to extrapolate		
the crossing points, $j^N_{c2}$, 
to the thermodynamic limit by employing a standard (BST) algorithm due
to Bulirsch and Stoer.~\cite{Henkel}
The method is explained in the Appendix. 
The results of this extrapolation procedure are shown in Fig. 4,
 where $j_{c2}$ is plotted as a function of $\Delta$.
The curve $j_{c2}$ separates a spin-gap (dimer) phase from a gapless phase
at small interchain couplings. As expected the value of $j_{c2}$ grows   
near the isotropic point.
(It should tend to $j=\infty$ 
at $\Delta = 1$ according to White and Affleck.~\cite{White96}) 

{\it Physical Probes.} Correlation functions
can also be used to determine the 
nature of different 
thermodynamic groundstates.
The spin rigidity, in particular,
can discriminate between phases that do and do not
have spin gaps. 
The stiffness of the nearest-neighbor spin-$1/2$ 
Heisenberg chain has been calculated
exactly via the Bethe ansatz.~\cite{Shastry} 
In the 
thermodynamic limit this solution yields 
\be
\frac{\rho_{11}}{J_1^{XY}} = 
\frac{\pi}{4} \sin \frac{\pi}{n} \frac{1}{\frac{\pi}{n} 
\left( \pi - \frac{\pi}{n} \right)}
\ee
for anisotropies
\be
\cos \frac{\pi}{n} = \frac{J_1^z}{J_1^{XY}} = \Delta,
\ee
where $n$ is a positive integer.
This yields
$\rho_{11}/J_1^{XY}=1/4$ 
in the isotropic case 
and $\rho_{11}/J_1^{XY}= 1/\pi$ 
 in the $XY$-case.
In Fig. 5 
we plot the exact Bethe ansatz stiffness~\cite{Shastry}
for the single chain and compare it
with the numerical diagonalization results for $N=8,12,16,20$
sites.
The spectrum is gapless from the $XY$-limit up to the 
isotropic Heisenberg case. Beyond this critical point the spectrum
acquires a gap due to the Ising anisotropy and the stiffness goes to zero. 
The stiffness remains positive beyond the transition point
for the small system sizes that we studied, however.
The anisotropy at which $\rho_{11} = J_1^{XY} / 4$
extrapolates nicely to the Heisenberg point, $\Delta = 1$,
in the thermodynamic limit, $N\rightarrow\infty$, however.

The zigzag ladder (or the chain with next-nearest neighbor interactions),
on the other hand,
is not solvable by the Bethe ansatz, and so other
methods become necessary to compute the spin rigidity.
With this purpose in mind,
exact diagonalization calculations of
the isotropic spin-1/2 Heisenberg
chain with next-nearest neighbors  interactions
where carried out first 
by Bonca et al.~\cite{Bonca} on small systems,
where the     diagonal component $\rho_{11}$ of the
stiffness  tensor was computed (see also ref. 31).
We have completed this study by calculating the remaining components
of the stiffness tensor (9), including cases with anisotropy.
In Fig. 6 we show the results for the various 
stiffnesses for the zigzag ladder in
the isotropic case as a function of $j$ 
for $N=16$, while the 
same set of results are shown for the $XY$-case ($\Delta = 0$) in Fig. 7.
In Figs. 6a and 7a we show 
the stiffnesses $\rho_{11}$, 
$\rho_{22}$ and $\rho_{12}$. For small $j$ 
we are in the limit  of strong interchain
coupling (small next-nearest neighbor interaction) and we
recover the previous results:~\cite{Bonca} 
$\rho_{11}>0$ and $\rho_{22}<0$. 
This is also found in the classical case.
In the opposite weak-coupling 
limit of very high $j$, $\rho_{11}$ remains negative,
but $\rho_{22}$ becomes positive 
(as in the classical case).  The latter is consistent with the
extreme limit of two decoupled antiferromagnetic chains ($J_1 = 0$).
Also, $\rho_{12}$ becomes non-zero in the
intermediate region where quantum fluctuations are stronger and where the
transition between weak-coupling and strong-coupling occurs.
Note that the classical analysis reveals
that the fact that $\rho_{12}$ becomes nonzero and 
the fact that
$\rho_{11}$ becomes negative are purely quantum effects.

In the initial exact diagonalization study for the isotropic case,   
Bonca et al. chose to  identify   the point at which the stiffness component
$\rho_{11}$ turns negative with a   transition point to a
quantum disordered phase with a spin gap.~\cite{Bonca}
Extrapolating to the
thermodynamic limit they estimated the transition point 
to occur at $j_c=0.270 \pm 0.005$.
We believe that it is a better idea  to look at the eigenvalues (11) of the
full stiffness tensor to determine possible phase transitions.~\cite{negative}
In Figs. 6b and 7b we plot $\rho_{+}$ for different system sizes. 
This stiffness is always positive for all $j$, 
while $\rho_{-}$ is always negative. 
For small $j$, $\rho_{+}$ is positive and of order unity. 
As $j$ grows $\rho_{+}$ becomes close to zero
at a value of $j$
that is close to the value where 
$\rho_{11}$ crosses zero,
 $j^N_{\rho 1}$.
As $j$ increases further 
$\rho_{+}$ again begins  to grow   appreciable
near the point at  which $\rho_{22}$ turns   positive, $j^N_{\rho 2}$.
A finite size analysis reveals that as the size $N$ grows
the first ``zero'' of $\rho_{+}$ (near $j = j^N_{\rho 1}$) 
occurs at smaller values of $j$ and extrapolates to
a value close to the dimer transition $j_{c1}$
determined by the level-crossing method. 
Similarly, we expect that
the second transition to a 
nonzero stiffness $\rho_{+}$ (near $j = j^N_{\rho 2}$) signals 
the transition to a gapless
regime that extends up to $j=\infty$,
and that it might therefore also signal a  decoupling transition. 
This happens, however,
both for $\Delta=0$ and for $\Delta=1$.
We believe that this is due to a finite-size effect
in the latter case as discussed above.
The fact that finite-size effects 
in the spin stiffness of a single chain 
become larger as the $XY$ anisotropy decreases (see Fig. 5)
supports this claim.

The level crossing at weak-coupling, $j^N_{c2}$, 
does not, however, correlate well with $j^N_{\rho 2}$. 
The crossings defined by $j^N_{\rho 2}$ appear at smaller values
of $j$ as compared with $j^N_{c2}$. 
Although both have the same trend, apparently
finite size effects are stronger in the calculation of the stiffness
than in the determination of the  level crossings.
We have therefore limited ourselves to the extrapolation of the level crossing
points $j^N_{c2}$.

In Fig. 8 we show the
behavior of the dimer correlation function $\chi_{11}$. 
For $j>j^N_{c1}$ the
dimer correlation function increases signaling the spontaneous dimerization
observed in the thermodynamic limit. 
The various peaks signal level crossings of the groundstate 
between near-degenerate states with  momentum $k=0$  and $k=\pi$
as $j$ varies. 
The Majumdar-Ghosh point at $j=1/2$ is special since 
the groundstate is doubly degenerate for all system sizes
in such case.
Beyond $j \sim 1$ the dimer
correlation function $\chi_{11}$ becomes small.
(The susceptibilities $\chi_{12}$ and $\chi_{22}$ are always quite small).

Finally, 
remnants of the spiral phase that exists in the  classical
limit, $S\rightarrow\infty$,
are clearly apparent in the momentum dependence of the
structure factor. In Fig. 9 we show the structure factor as 
a function of momentum
for several values of $j$ in the Heisenberg case at $\Delta=1$. 
We see that the location of the maximum shifts from
$k_{max}=\pi$ (for $j \leq 0.5$) to 
a value $\pi/2 < k_{max} < \pi$ when $j>0.5$,
thereby signaling the spiral phase. 
The results are similar for $\Delta=0$.

\section{Conclusions}
The $S = 1/2$ antiferromagnetic zig-zag ladder is a difficult problem
to solve due to the intrinsic frustration and to the criticality
displayed by both the strong and the weak-coupling limits.
The weak-coupling limit  is particularly difficult in the absence of
anisotropy, in which case antiferromagnetic correlations 
diverge   exponentially.~\cite{White96} 
This renders any numerical study of finite systems hard.
 
In this paper,
we have performed
 an analysis of the exact properties of
such finite systems, looking at various correlation functions and
the structure of the spectrum both in the isotropic and the
anisotropic cases. 
We have looked at the spectrum and have computed the
spin stiffness of the
zig-zag ladder, and have thereby found evidence 
for a gapless regime at weak coupling  that survives
the  thermodynamic limit
in the case of	$XY$ anisotropy. 
However, 
the isotropic and the anisotropic
cases look qualitatively similar.
We believe that this is    
due to a strong  finite-size effect in the former case.	
This claim is supported by the increase of finite-size
effects in the stiffness 
of a    single  $S = 1/2$ antiferromagnetic chain
with  decreasing $XY$ anisotropy,
as shown  in Fig. 5. 

It was previously shown~\cite{Okamoto} that the 
dimer transition can be accurately determined
in relatively small systems by studying the level crossing of the first
and second excited states after extrapolating to the 
thermodynamic limit. 
We have used
a similar criterion to detect a possible transition from the dimer phase to
a gapless phase at weak-inter chain coupling. Using standard extrapolation
techniques (see Appendix) we have constructed a phase diagram
(see Fig. 4) in this regime that is in agreement
with a recent proposal for a gapless spiral phase 
in the presence of $XY$ anisotropy.~\cite{Nersesyan}

Also, we expect the spin stiffness
$\rho_+$ to be a good measure of the nature of the 
spectrum.  In particular, 
it can be used as an order parameter to distinguish 
gapless from spin-gap phases. 
A positive stiffness indicates a gapless excitation spectrum and a null
stiffness indicates  a net  spin-gap.  
The antiferromagnetic zig-zag for spin $S = 1/2$
 showed an appreciable positive stiffness in the limit
of strong interchain coupling (similar to 
the single chain case),
a very small yet positive stiffness in the intermediate (spin-gap) 
regime,
and then  again an appreciable positive stiffness in 
the limit of weak interchain coupling 
(similar  to two decoupled  chains).
This was true for all values of 
the ($XY$) anisotropy, $\Delta\leq 1$.
It is known from previous work,~\cite{White96} however,
that a spin gap is always expected to
be present in the isotropic case at weak coupling.
The stiffness should therefore  remain zero in the thermodynamic limit
at $\Delta = 1$ in the weak-coupling regime.
We believe that the  discrepancy between this expectation
and our results is a strong
finite size effect in the isotropic case that 
is due to the exponentially diverging 
antiferromagnetic correlations.~\cite{White96}
Clearly, larger systems need to be studied.

\bf
Acknowledgments
\rm

We have benefited from helpful discussions and comments
from Didier Poilblanc, Alexander Nersesyan, Fabian Essler and Alexander
Gogolin.  
We are also indebted to Ed Rezayi for suggesting the
spectral analysis.

\appendix
\section*{}

{\it Exact Diagonalization.}
The size of the Hilbert space under consideration can be considerably 
reduced using the symmetries of the problem. 
The Hamiltonian commutes with the total spin operator $S^{z}_{T}$,
with the translation operator $T$, 
the spin flip operator $R$, 
and the space reflection operator $L$ $(i\rightarrow N+1-i, i=1,\ldots,N)$. 
In the absence of anisotropies the Hamiltonian also commutes with the total 
spin operator ${(\vec{S}_{T})}^2$ and the energy levels come in spin 
multiplets.

The operator $R$ commutes with $T$ but, although $L$ commutes with $R$, it 
does not commute with $T$, since $LT={T}^{-1}L$.

The action of the local operators $S^{\pm}_{i}$ and $S^{z}_{i}$ is simply
given in the direct product basis $|m^{z}_{1}>\cdots|m^{z}_{N}>$, which are 
eigenvectors of $S^{z}_{T}$. 
In general, these states are not eigenvectors of the additional symmetries.
We consider first the translations, obtaining the classes of states which 
are closed under them. 
One starts with a state $|a>$ and applies $N_{a}$ times the translation 
operator $T$ until one finds $T^{N_{a}}|a>=|a>$, where $N_{a}$ is 
necessarily a divisor of $N$. 
The state $|a>$ is the representative of this class. In combinatorial 
theory this is called a necklace and its periodic part, of length $N_{a}$, 
a Lyndon word. 
The other classes are formed proceeding in the same manner starting with 
other states, not already used, until all the states have been exhausted. 
In each class, since $T^{N_{a}}=\hat{1}$, the possible eigenvectors of $T$ 
are $t_{N_a}(k)=e^{\frac{i2\pi k}{N_a}}$, with $k=0,\ldots,N_a-1$, 
corresponding to the momentum $p_k=\frac{2\pi k}{N_a}$.
The operators 
$P_N(k)=\frac{1}{N}\sum_{i=0}^{i=N-1}(\frac{T}{t_{N}(k)})^{i}$ 
are projectors, i.e, they satisfy 
$P_N(k)P_N(k^\prime)=\delta_{k,k^\prime}P_N(k)$. 
They also have the property $TP_N(k)=t_{N}(k)P_N(k)$.
The projector $P_N(k)$ acting on the states of a class with $N_a$ elements 
reduces to $\frac{N}{N_a}P_{N_a}(k)$, if $t_{N}(k)$ is one of the 
$t_{N_a}(k)$ and gives zero otherwise.
The normalized eigenvector of momentum $p_k$ formed from the class of the 
state $|a>$ is given by $|k,a>=\sqrt{N_a}P_{N_a}(k)|a>$.
Since the Hamiltonian commutes with $S^{z}_{T}$ and $T$, the subspaces of
fixed magnetization and momentum are invariant subspaces and it is 
important to consider each of these subspaces separately. A general state 
is written as a linear combination of the states $|k,a>$ corresponding to 
that magnetization. The state $|k,a>$ is represented by $|a>$. This allows 
us to reduce the size of the basis to the number of representatives.

The spin reflection changes the sign of the magnetization. Therefore it is 
relevant only for the classification of the states when the magnetization 
is zero. IF $R|a>$ is not in the class of $|a>$ both eigenvectors with 
$r=1$ and $r=-1$ can be formed. 
IF $R|a>$ is in the class of $|a>$, it must be obtained from this state by 
$\frac{N_a}{2}$ translations (implying that $N_a$ must be even), since for 
half-integer spin it is linearly independent of $|a>$. 
The state $|k,a>$ is an eigenvalue of $R$ with $r=1$ for $k$ 
even and $r=-1$ for $k$ odd.

The space reflection operator reverses the momentum, since one has 
$LP_N(k)=P_N(-k)L$. The state $L|a>$ can be in a different class or in the 
same class. The states $|k,a>$ and $|-k,a>$ are linearly independent, 
except for $k=0$, with eigenvalue $l=1$, or $k=\frac{N_a}{2}$, for $N_a$ 
even, with eigenvalue $l=-1$. 
As a result, one sees that, for zero magnetization and $t=\pm 1$, it is 
possible to use all the operators $T,R$ and $L$ in the construction of 
the states. This allows a further reduction of the size of the subspace 
under consideration. Since some classes have a definite eigenvalue 
of $R$ or $L$ they are simply excluded if their values are not those of the 
state which we are studying.

Since the complete Hamiltonian commutes with the operator $T$ only 
transitions to states with the same momentum are allowed, even if separate 
terms in the Hamiltonian allow them. 
If $H|a>=\sum_b\alpha T^{p(a,b)}|b>$, where $p(a,b)$ is an integer and 
$|b>$ is the representative of a class, one finally finds 
$ <k,b|H|k,a>=\sum_b^\prime\alpha\sqrt{\frac{N_a}{N_b}}\left(t_{N_a}(k)\right)
^{p(a,b)}$, where the summation is restricted to the allowed transitions. 

The modified Lanczos method~\cite{Gagliano} is very useful to obtain the 
ground state of an Hamiltonian. Restricting our attention to a subspace 
of fixed magnetization and momentum we will find the ground state of the 
block Hamiltonian in that subspace.
To obtain an approximate ground state we choose a trial state $|\psi_0>$ 
that can not be orthogonal to the true ground-state. 
We define a state $|\psi_1>$ as~\cite{Gagliano}
\be
|\psi_1> = \frac{ \hat{H} |\psi_0> - 
<H> |\psi_0>}{\left( <H^2>-<H>^2 \right)^{1/2}}
\ee
where $<\psi_1 |\psi_1 >=1$, $<\psi_1 | \psi_0>=0$ and
$<H^n>=<\psi_0 | \hat{H}^n |\psi_0>$. Defining a matrix of the Hamiltonian
in the basis $|\psi_0>,|\psi_1>$ we can diagonalize it~\cite{Gagliano} 
obtaining a next order approximation for the energy $\tilde{\epsilon}_0$, 
and ground state $|\tilde{\psi}_0>$, with
\bea
\tilde{\epsilon}_0 & = & <H> + b \alpha \\
|\tilde{\psi}_0> & = & \frac{|\psi_0> + \alpha |\psi_1> }{\left( 1+\alpha^2
\right)^{1/2}}
\eea
where
\bea
b & = & \left( <H^2>-<H>^2 \right)^{1/2} \\
\alpha & = & f-\left( f^2 +1 \right)^{1/2}
\eea
and
\be
f = \frac{<H^3> - 3<H> <H^2> + 2 <H>^3}{2 \left(<H^2> - <H>^2 \right)^{3/2}}.
\ee
Taking $|\tilde{\psi}_0>$ as the new $|\psi_0>$ we can iterate the method
to obtain a better estimate for the energy and the ground state.

Since the ground state and the first excited of the system have different 
momenta they are both ground states of different subspaces, and we apply 
the modified Lanczos method to these subspaces. If the system does not 
have anisotropies one can alternatively look for the component of the spin 
multiplet of the first excited state with magnetization $1$.

{\it Extrapolation.}
The results for the infinite system can be estimated using standard 
extrapolation methods like  
the BST method.~\cite{Henkel} 
In the BST algorithm we look for the limit of a sequence of the type
$T(h_j)=T +a_1 h_j^{\omega} + a_2 h_j^{2 \omega}+ ... $, $j=0,\cdots,N_p-1$ 
($N_p$ being the number of data points), where $h_j$ is a sequence converging 
to zero as $j\rightarrow\infty$, corresponding to different system sizes $N_j$. 
Typically $h_j=\frac{1}{N_j}$, with $N_j=a+bj$, for some values of $a,b$.
The value of the $m^{th}$ iteration for the sequence is obtained from
\be
T_m^{(j_m)}=T_{m-1}^{(j_m+1)}+\frac{T_{m-1}^{(j_m+1)}-T_{m-1}^{(j_m)}}
{\left(\frac{h_{j_m}}{h_{j_m+m}}\right)^{\omega}\left(1-\frac{T_{m-1}
^{(j_m+1)}-T_{m-1}^{(j_m)}}{T_{m-1}^{(j_m+1)}-T_{m-2}^{(j_m+1)}}\right)-1}
\ee
with $j_m=0,\cdots,M(m)$ and where 
$M(m)$ is the number of values of $T_m^{(j_m)}$ 
at each iteration. 
It decreases by one, at each iteration, from $M(0)=N_p$ to $M(N_p-1)=1$, when 
the iteration process is fulfilled. 
As initial values one defines $T_{-1}^{(j)}=0$ and $T_0^{(j)}=T(h_j)$. 
The extrapolated value is $T_{N_p-1}^{(0)}$ and the estimated error is 
$\epsilon=|T_{N_p-2}^{(1)}-T_{N_p-2}^{(0)}|$.
Finally, $\omega$ is a free parameter which is adjusted such that the estimate 
of the error is a minimum.

\end{multicols}

\newpage

\vspace{\baselineskip}
Table I- Lowest energy levels for $N=16$. The states are represented
by their $S_z$ values and momenta $(S_z;k)$. When more than one
state is represented this means they are degenerate. When the momenta
are not $0$, $\pi$ or $\pi/2$ the momentum of the state is represented
by an integer, $n$, such that $k=\frac{2 \pi}{N}n$.

\vspace{\baselineskip}

$
\begin{array}{llll}
 j & \Delta=0 & \Delta=0.5 & \Delta=1 \\
  &      &    &    \\
  & (2;\pi) & (0;0) & (0,1,2;0,\pi) \\
  & (0;\pi) & (0;\pi) & (0;\pi/2) \\
  & (1;\pi/2) & (1;\pi/2) & (0,1;\pi/2) \\
10 & (0;0) & (0;0) & (0;0) \\
   &       &       &       \\
   & (2;\pi) & (0;0) & (0;\pi/2) \\
   & (0;\pi) & (0;\pi) & (0;\pi) \\
   & (1;\pi/2) & (1;\pi/2) & (0,1;\pi/2) \\
2  & (0;0) & (0;0) & (0;0) \\
   &       &       &       \\
   & (2;\pi) & (0;\pi/2) & (0;\pi/2) \\
   & (0;\pi) & (1;\pi/2) & (0,1;\pi/2) \\
   & (1;\pi/2) & (0;\pi) & (0;\pi) \\
1.5 & (0;0) & (0;0) & (0;0) \\
    &       &       &       \\
    & (0;\pi) & (0;\pi) &       \\
    & (1;\pi/2) & (1;\pi/2) & (0,1;\pi/2) \\
    & (0;\pi) & (0;\pi) & (0;\pi) \\
1   & (0;0) & (0;0) & (0;0) \\
    &       &       &       \\
    & (1;3) & (1;3) & (0,1;\pi) \\
    & (0;\pi) & (0;\pi) & (0,1;3) \\
    & (0;0) & (0;0) & (0;0) \\
0.75 & (0;\pi) & (0;\pi) & (0;\pi) \\
     &         &         &         \\
     & (0;0) & (1;1) & (0,1;1) \\
     & (1;\pi) & (1;\pi) & (0;0) \\
     & (0;\pi) & (0;\pi) & (0,1;\pi) \\
0.5  & (0;0,\pi) & (0;0,\pi) & (0;0,\pi) \\
     &           &           &           \\
     & (0;\pi)  & (0;\pi) & (0,1;1) \\
     & (0;\pi) & (0;\pi) & (0,1;\pi) \\
     & (1;\pi) & (1;\pi) & (0;\pi) \\
0.25 & (0;0) & (0;0) & (0;0) \\
     &       &       &       \\
     & (0;\pi) & (0;\pi) & (0,1;1) \\
     & (0;\pi) & (0;\pi) & (0;\pi) \\
     & (1;\pi) & (1;\pi) & (0,1;\pi) \\
0.1  & (0;0) & (0;0) & (0;0) \\
     &       &       &       
\end{array}
$

\newpage
\bf
\begin{center}
Figure Captions
\end{center}
\rm

\vspace{\baselineskip}
\noindent
Fig. 1- Diagram of the zig-zag ladder. The nearest-neighbor interaction is
parameterized by $J_1$ and the next-nearest-neighbor interaction is parameterized
by $J_2$.

\vspace{\baselineskip}
\noindent
Fig. 2- Classical stiffnesses $\rho_{11} / (J_1 S^2)$, 
$\rho_{22} / (J_1 S^2)$ and 
$\rho_s / (J_1 S^2)$ as a function of $j = J_2 /J_1$. 

\vspace{\baselineskip}
\noindent
Fig. 3- Crossing points of the first excited state as a function of $N$
for different values of $\Delta$ at weak coupling.

\vspace{\baselineskip}
\noindent
Fig. 4- Phase diagram in the regime of weak interchain coupling. 
The critical coupling parameter $j_{c2}$ is
obtained from the crossing of the first excited state with the second excited
state after extrapolating to the thermodynamic limit (see the Appendix). 

\vspace{\baselineskip}
\noindent
Fig. 5- Comparison of the exact Bethe ansatz result for the spin stiffness
of a single chain to the numerical results obtained from diagonalizing
small systems as a function of $\Delta$.

\vspace{\baselineskip}
\noindent
Fig. 6- (a) Stiffnesses $\rho_{11}$, $\rho_{12}$ and $\rho_{22}$ as a function
of $j$ for $\Delta=1$ and $N=16$. (b) $\rho_+$ as a function of $j$ for various
system sizes for $\Delta=1$.

\vspace{\baselineskip}
\noindent
Fig. 7- (a) Stiffnesses $\rho_{11}$, $\rho_{12}$ and $\rho_{22}$ as a function
of $j$ for $\Delta=0$ and $N=16$. (b) $\rho_+$ as a function of $j$ for various
system sizes for $\Delta=0$.

\vspace{\baselineskip}
\noindent
Fig. 8- Dimer correlation function
for $N=16$ and $\Delta=1$ and $\Delta=0$.

\vspace{\baselineskip}
\noindent
Fig. 9- Structure factor, $S(q)$, for $N=16$ and $\Delta=1$ for various
coupling strengths.
The peak is located  at $q=\pi$ in the single-chain limit, $j=0$
but this momentum tends to $q=\pi/2$ as $j$ increases.
This indicates the presence of important spiral correlations in the
weakly-coupled $S = 1/2$  zig-zag antiferromagnet.


\begin{references}
\bibitem{extrap}
An alternative way to determine a
transition into a spin-gap phase is to extrapolate 
the values of the gaps for spin excitations
in the two regimes to the thermodynamic limit.
This is a difficult numerical problem because it is hard to distinguish
between a very small gap 
and a null gap.
\bibitem{negative}
The stiffness can be negative 
for finite systems
even though it is expected to be either
zero or positive  for infinite systems. 
In the thermodynamic limit, 
a negative stiffness means that the system is unstable.
In the case of a finite system it may mean that the true
groundstate is incommensurate with the system 
size considered (see ref. 31).
\end{references}
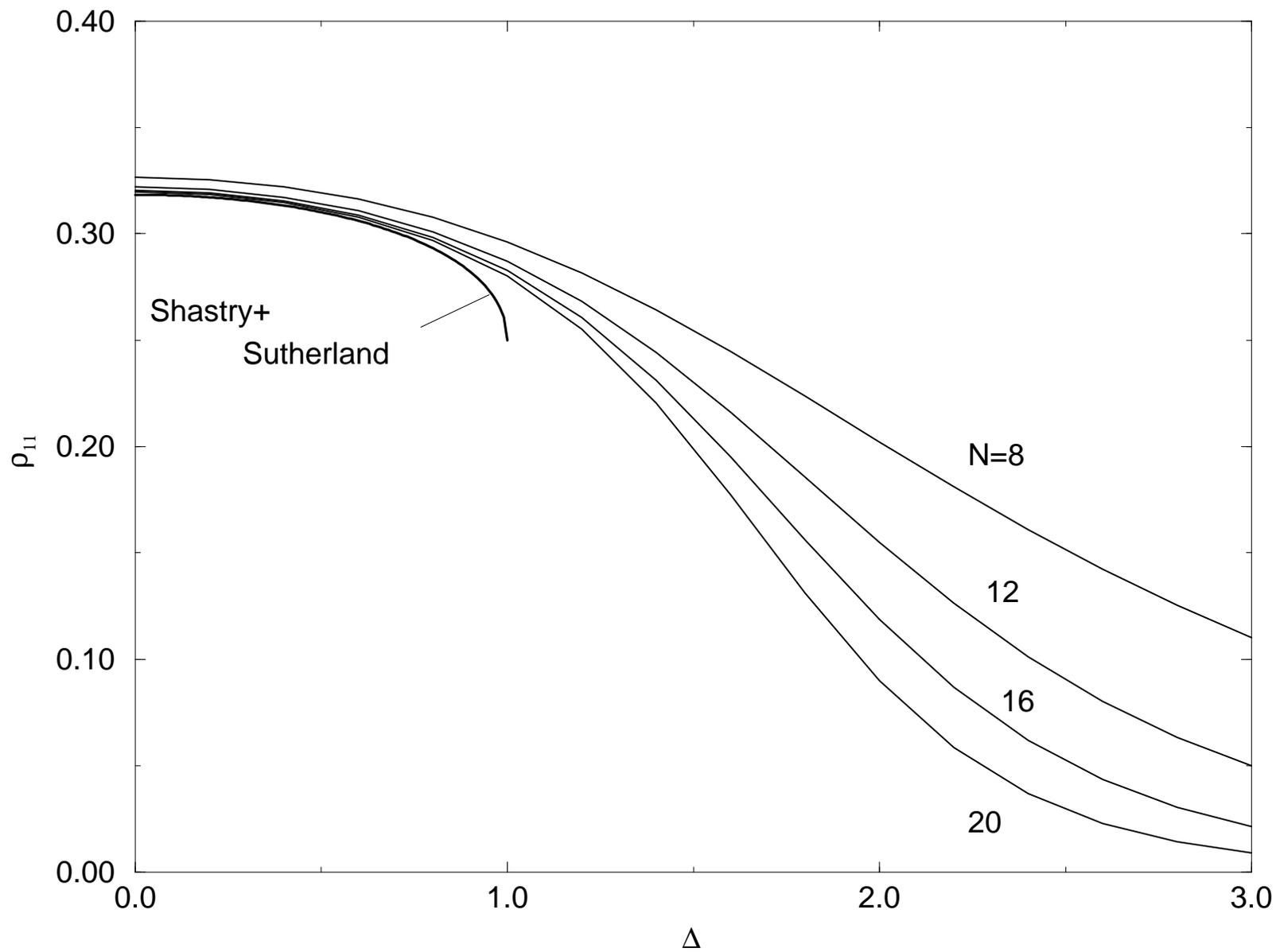
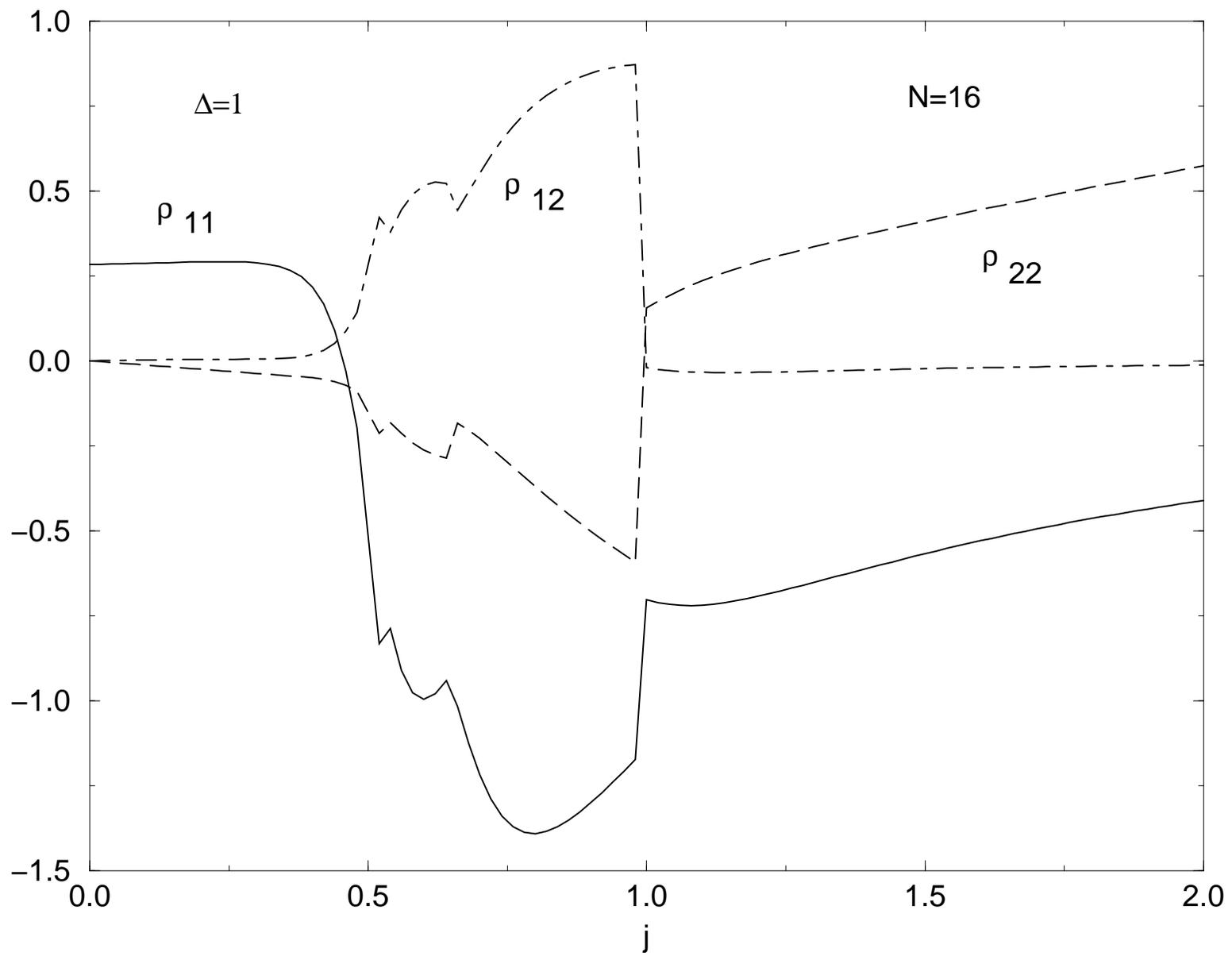
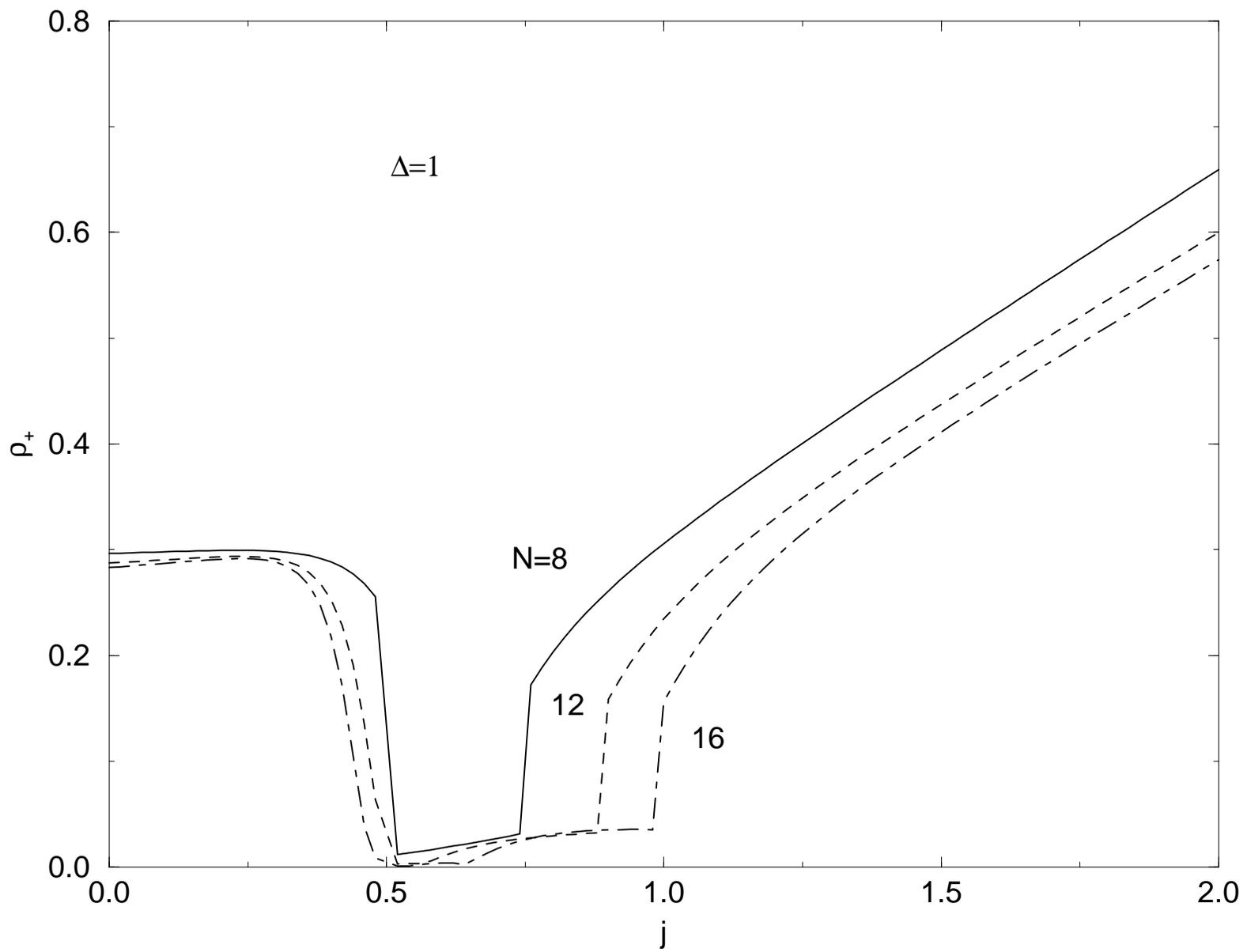
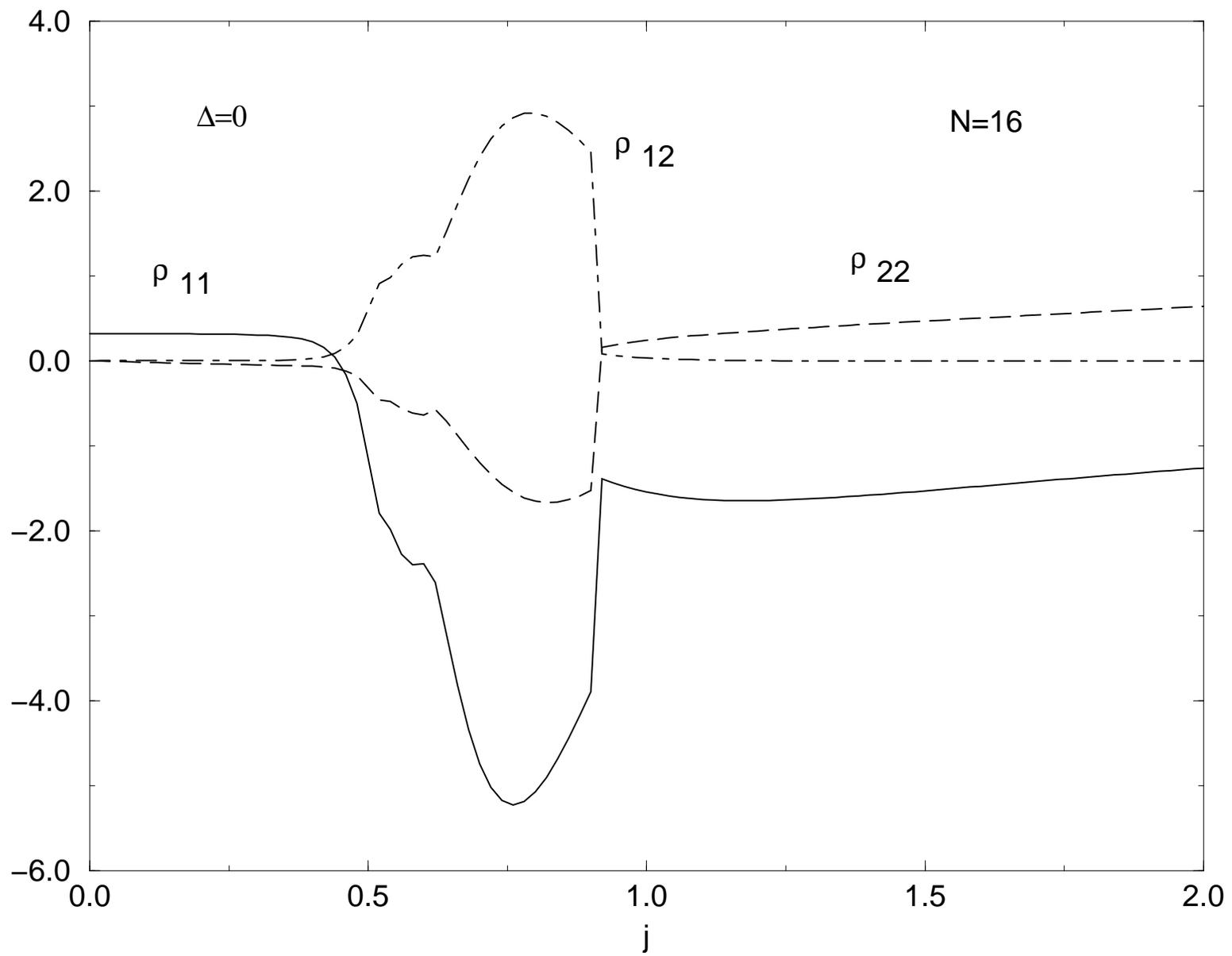
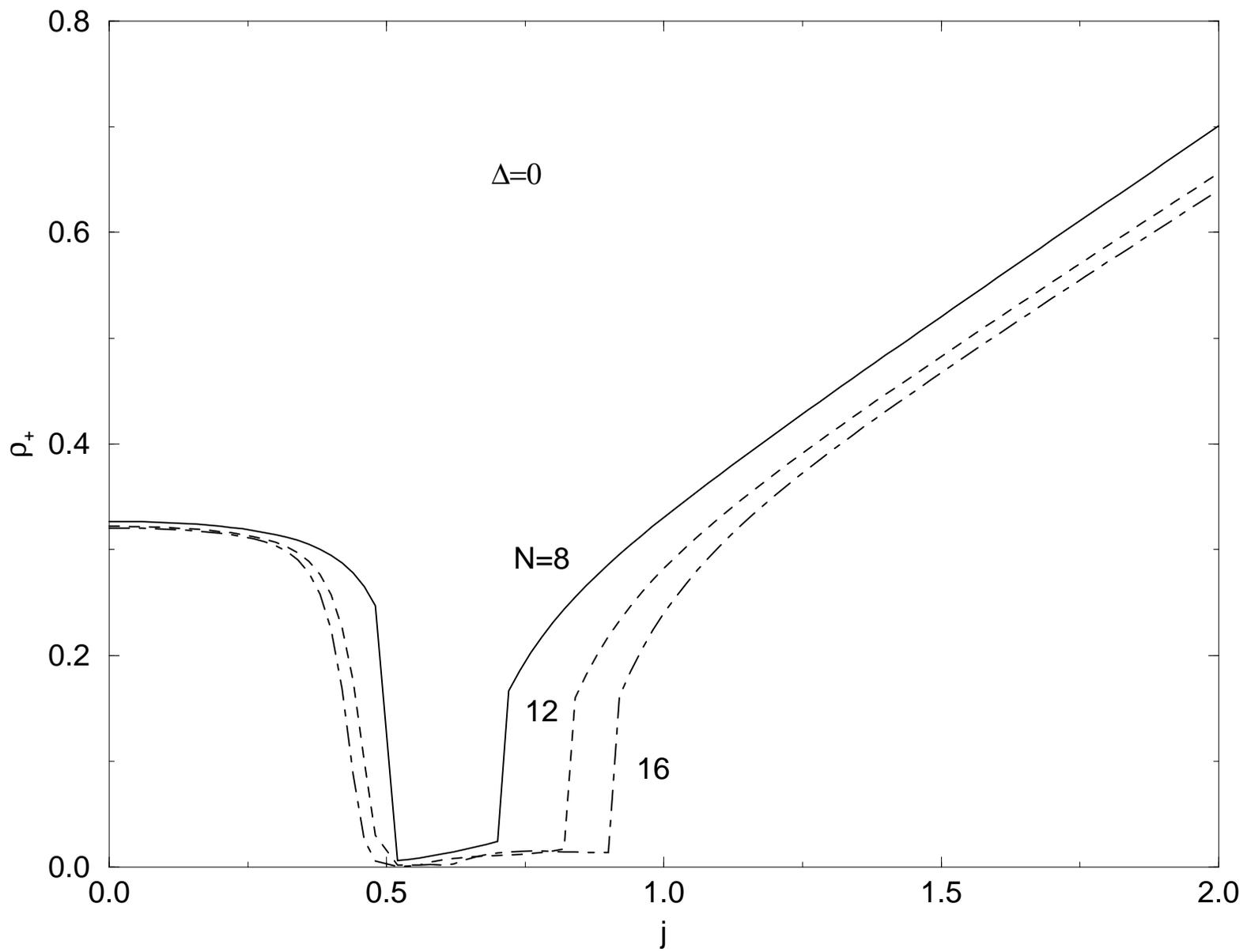
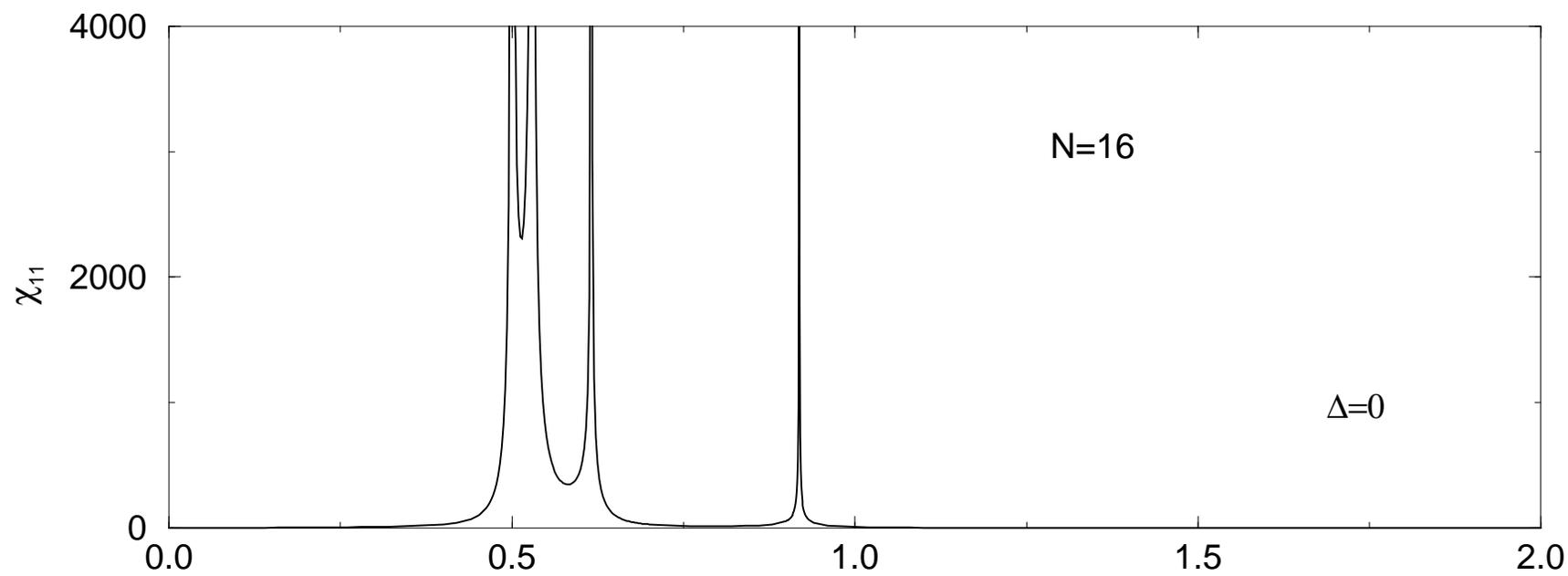
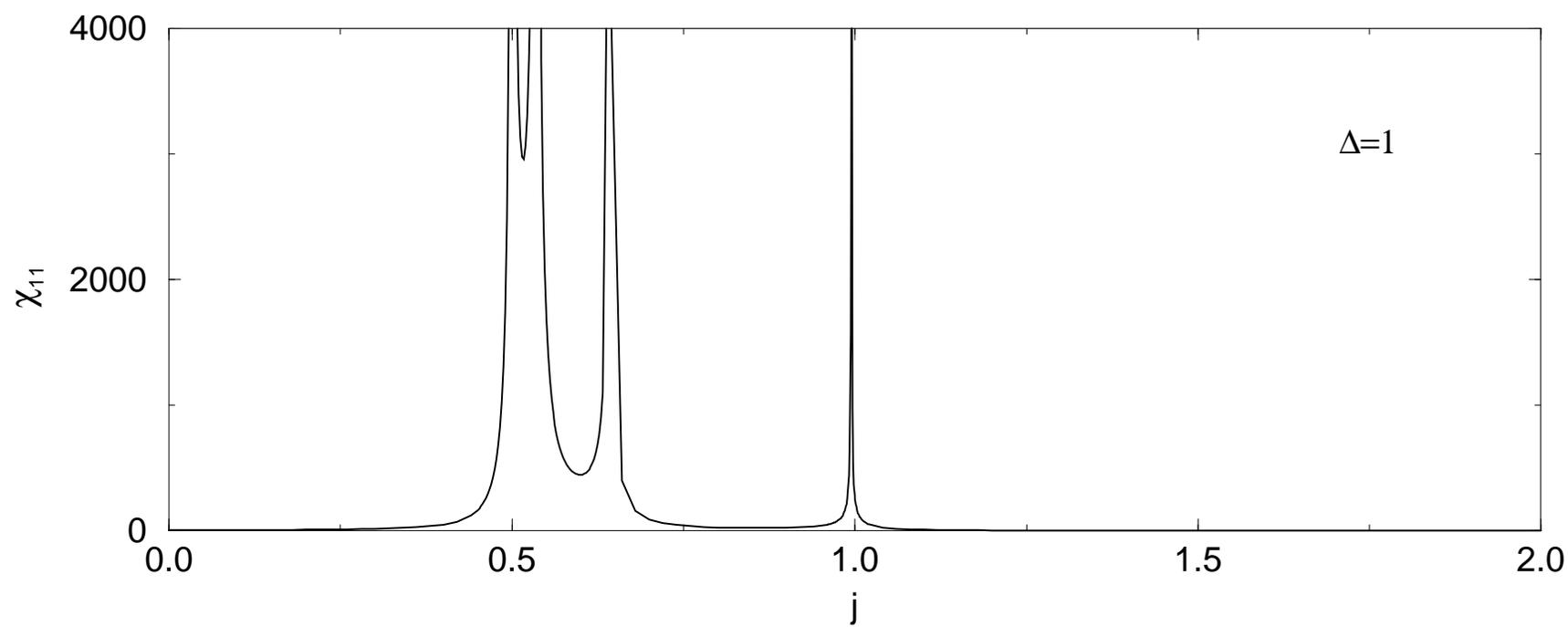
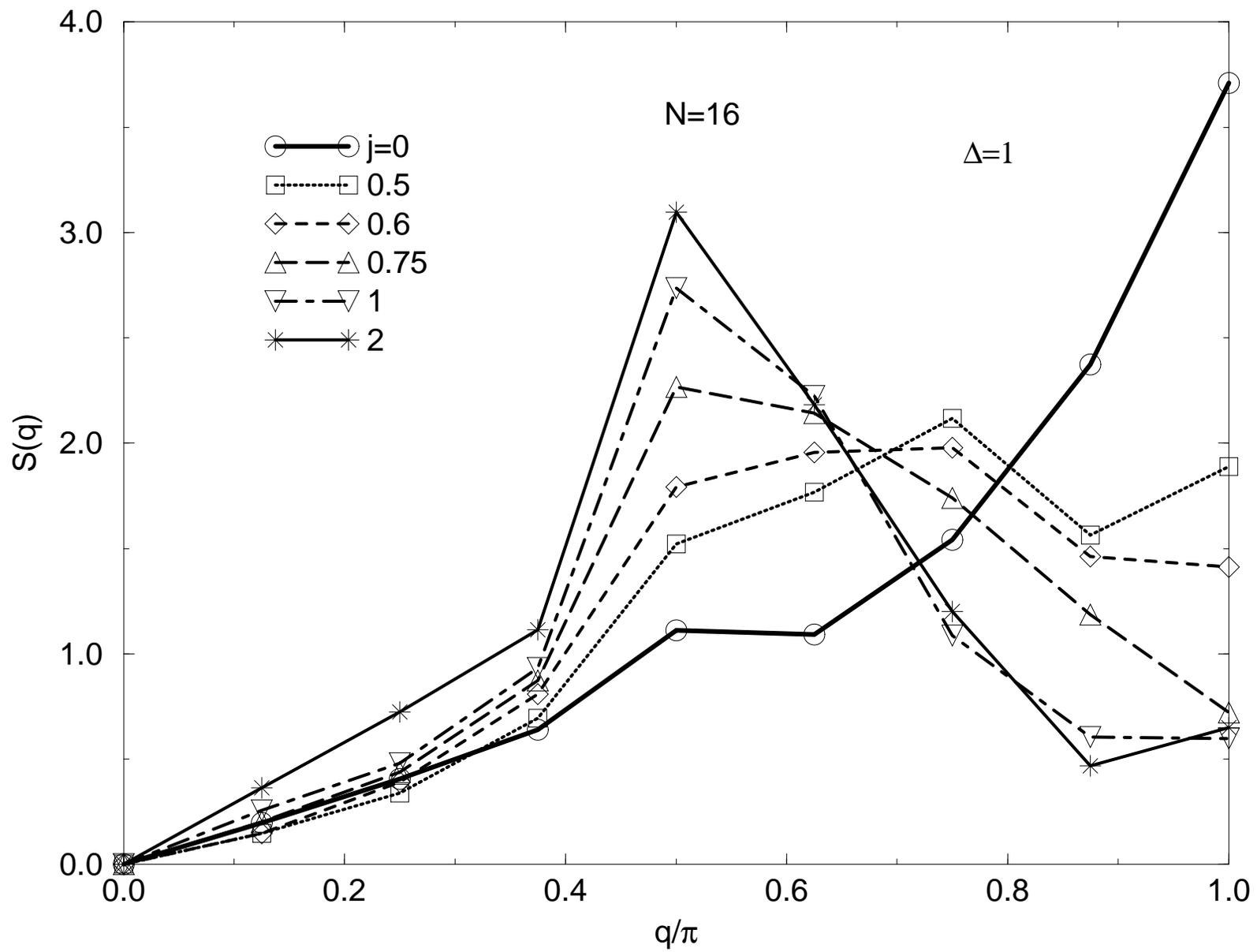
\end{document}